\documentclass[eqsecnum,aps,12pt]{revtex4}
\usepackage{graphicx}
\usepackage{dcolumn}
\usepackage{bm}
\usepackage{color}

\newcommand{\bq}{{\bf q}}
\newcommand{\bv}{{\bf v}}

\newcommand{\br}{{\bf r}}
\newcommand{\bqp}{{\bf q}_\perp}
\newcommand{\qpa}{q_\parallel}
\newcommand{\brp}{{\bf r}_\perp}
\newcommand{\rpa}{r_\parallel}
\newcommand{\Rp}{R_\perp}
\newcommand{\Rpa}{R_\parallel}

\newcommand{\bR}{{\bf R}}
\newcommand{\bRp}{{\bf R}_\perp}

\newcommand{\bff}{{\bf f}}

\newcommand{\hxp}{\hat{x}_\parallel}

\newcommand{\hrp}{\hat{r}_\perp}

\newcommand{\beq}{\begin{equation}}
\newcommand{\eeq}{\end{equation}}
\newcommand{\beqn}{\begin{eqnarray}}
\newcommand{\eeqn}{\end{eqnarray}}
\newcommand{\pp}{\partial}

\newcommand{\rp}{{r_\perp}}

\def\dry{polar ordered dry active fluid}

\begin{document}

\begin{center}
{\large {\bf  Giant number fluctuations in dry active polar fluids: A shocking analogy with lightning  rods}}
\vskip .75cm
   {\bf  John Toner }
\vskip.5cm
   { Institute of Theoretical Science\\ and
Department of Physics\\ University of Oregon, Eugene, OR
97403-5203}
\end{center}

\section*{Abstract}

The hydrodynamic equations of dry active polar fluids (i.e., moving flocks without momentum conservation) are shown to imply giant number fluctuations. Specifically,  the rms  fluctuations $\sqrt {<(\delta N)^2>}$ of the number $N$ of active particles  in a region containing a mean number of active  particles $<N>$ scales according to the law $\sqrt {<(\delta N)^2>} = K'<N>^{\phi(d)}$
with
$\phi(d)=\frac{7}{10}+\frac{1}{5d}$ in $d\le4$ spatial dimensions. This is much larger the
``law of large numbers" scaling $\sqrt {<(\delta N)^2>} = K\sqrt{<N>}$
found in most equilibrium and non-equilibrium systems.
In further contrast to most other systems,
the coefficient
$K'$
also depends singularly on the shape of the box in which one counts the particles, vanishing in the limit of very thin boxes. These fluctuations arise {\it not} from large density fluctuations - indeed, the density fluctuations in \dry s are not in general particularly large -  but from  long ranged
spatial correlations between those fluctuations. These are shown to be  closely related in two spatial dimensions to the electrostatic potential near a sharp upward pointing conducting wedge of opening angle ${3\pi\over8}=67.5^\circ$, and in three dimensions to  the electrostatic potential near a sharp upward pointing charged cone of opening angle $37.16^\circ$. This very precise prediction can be stringently  tested   by alternative box counting experiments that directly  measure this density-density correlation function.
% \end{abstract} 

 \section{Introduction}
 
Non-equilibrium systems in general, and active matter in particular, can exhibit many novel behaviors impossible in equilibrium systems. One of the most striking examples is the existence of long-ranged order associated with a broken continuous symmetry in two dimensions (2D) -- a phenomenon forbidden in equilibrium systems by the Mermin-Wagner theorem \cite{MW}. Collective motion, or ``flocking'', can therefore exist in active matter, even in two dimensions \cite{TT1,TT2, TT3,TT4, NL}.  

Another striking phenomenon that can occur in active matter is Giant number fluctuations (GNF).  These were first predicted to occur for dry (i.e., non-momentum conserving) {\it apolar} active fluids (also known as ``active nematics")\cite{actnemsub, act nem}. It was later noted that these should also occur in dry  {\it polar} active fluids (also known as  ``ferromagnetic flocks")\cite{Chate+Giann}).

The phenomenon of Giant number fluctuations can be detected simply by counting, as follows:

Within a large \dry, identify some smaller sub-volume, which I'll hereafter call the ``counting box", that is still large enough to contain an enormous number $N$ of particles. Count the number of particles in it. Repeat this count in the same volume many times, as the system evolves. Once enough statistics have been collected, determine the mean number $<N>$ of particles, and its rms fluctuation $\sqrt{<(\delta N)^2>}$, where $\delta N\equiv N-<N>$. Now repeat this process with a sequence of progressively larger boxes. As one does so, both the mean number of particles$<N>$, and its variance $\sqrt{<(\delta N)^2>}$, will increase. 

In virtually all equilibrium systems\cite{BEC}, and most non-equilibrium systems, the result of such an analysis will be the so-called ``law of large numbers": $\sqrt{<(\delta N)^2>}\propto\sqrt{<N>}$. But in \dry, I find that these fluctuations are far larger.

Specifically, I derive the existence of these Giant number fluctuations directly from the hydrodynamic equations\cite{TT1,TT2,TT3,TT4,NL} of \dry s. Making a plausible conjecture about the scaling laws   implied by those hydrodynamic equations, I find that the rms number fluctuations $\sqrt {<(\delta N)^2>}$ of the number $N$ of active particles  in a region containing a mean number of active  particles $<N>$ scales according to the law \begin{eqnarray}
\sqrt {<(\delta N)^2>} = K'<N>^{\phi(d)}
\label{Nfluc}
\end{eqnarray}
with
\begin{eqnarray}
\phi(d)=\frac{7}{10}+\frac{1}{5d}  
\label{phi}
\end{eqnarray}
 in $d\le4$ spatial dimensions. Since $\phi(d)>1/2$ in all $d<4$ \cite{d>4},this is much larger the
``law of large numbers" scaling $\sqrt {<(\delta N)^2>} = K\sqrt{<N>}$
found in most equilibrium and non-equilibrium systems.

Even stranger is the fact that, in further contrast to most other systems,
the coefficient
$K'$
also depends singularly on the {\it shape} of the box in which one counts the particles, vanishing in the limit of very thin boxes. 

Considering, for example, a three dimensional system in which I count particles in a box that I call a ``needle shaped" counting volume. I define this as a long thin cylinder with its axis along the direction of the mean average velocity $<\bv>$ of the \dry, with aspect ratio $\beta\equiv{L_\parallel\over L_\perp}\gg1$, where $\L_\parallel$ and $\L_\perp$ are respectively the length of the cylinder axis, and its radius respectively. For such a shape, I find
\begin{eqnarray}
K'=\propto\beta^{-23/30} \,.
\label{K'3d}
\end{eqnarray}

In contrast, for a ``pancake shaped" box, which I  define as a squat cylinder again with its axis along the direction of the mean average velocity $<\bv>$ of the \dry, a direction I will hereafter refer to as $\hxp$, but now with aspect ratio $\beta\equiv{L_\parallel\over L_\perp}\ll1$, where $\L_\parallel$ and $\L_\perp$ are respectively the length of the cylinder axis, and its radius respectively. For this shape, I find

\begin{eqnarray}
K'\propto\beta^{8/15} \,.
\label{K'pan3}
\eeqn

Similar results hold in $d=2$. Here, I consider a rectangular counting box aligned with two of its edges parallel to $\hxp$. In the ``needle" limit, this will be the long axis, while for the ``pancake" limit (which in $d=2$ is just the needle rotated by $90$
 degrees), it will be the short axis. 
 
 Continuing to define $\beta\equiv{L_\parallel\over L_\perp}$ in all cases, for the needle case $\beta\gg1$, I find
 \begin{eqnarray}
K'\propto\beta^{-1/5} \,,
\label{K'2dneed}
\end{eqnarray}
while for the pancake case $\beta\ll1$, I find
 \begin{eqnarray}
K'\propto\beta^{1/5} \,.
\label{Nfluc2dpan}
\end{eqnarray}
This has the appealingly symmetric feature that the giant number fluctuations are comparable if obtained from a long cylindrical counting box aligned with its long axis along $\hxp$ as in the same box rotated by $90^\circ$ to align with its long axis perpendicular to $\hxp$.

These fluctuations arise {\it not} from large density fluctuations - indeed, the density fluctuations in \dry s  are not in general particularly large -  but from  long ranged
spatial correlations between those fluctuations. I find that in two spatial dimensions, these are related, by a simple anisotropic spatial rescaling, in the upper half plane, from the electrostatic potential near a sharp upward pointing conducting wedge of opening angle ${3\pi\over8}=67.5^\circ$. Likewise, in  three dimensions, an identical rescaling connects  density correlations in the upper half space to  the electrostatic potential near a sharp upward pointing charged cone of opening angle $37.16^\circ$.

Here, by upper, I mean in the direction of the mean velocity $<\bv>\equiv v_0\hat{{\bf x}}_\parallel$ of the \dry, which is by definition non-zero in the ordered state, to which all these results are limited. 

Specifically, I find
\begin{eqnarray}
 C_{\rho} ({\bf r}) = r^{-\alpha(d)} G_d(\theta_{{\bf r} })
\label{Crhorscale}
\end{eqnarray}
where $r$ is the magnitude of  ${\bf r}$ $ ( r=|{\bf r}| )$, 
$\theta_{{\bf r}}$ is the angle between  ${\bf r}$ and the direction of
mean flock motion
$\hat{{\bf x}}_\parallel$ (defined in $d=2$ to run from $-\pi$ to $\pi$, and in $d=3$ to run between $0$ and $\pi$), and 
\beq
\alpha(d)={3d-2\over5} \,.
\label{alpha1}
\eeq
The function $G_d(\theta_{{\bf r}})$ is given by 
\begin{eqnarray}
G_d(\theta_{{\bf r}}) \equiv {\Upsilon_d(\theta_{{\bf R}})\over\left[{c^2\over (\gamma-v_2)^2}\cos^2\theta+\sin^2\theta\right]^{\alpha(d)/2}} \,,
\label{Gdef}
\end{eqnarray}
where 
\beqn
\theta_\bR=\tan^{-1}\left({|\gamma-v_2|\over c}\tan(\theta_\br)\right) \,,
\label{thetaR}
\eeqn with $c$, $\gamma$, and $v_2$ system dependent parameters defined in section II,
and
the function $\Upsilon_d$ given by 
\beqn
\Upsilon_2(\theta_\bR)&=&\left\{
\begin{array}{ll}
B_{2}\cos(4\theta_\bR/5)\,\,\,\,\,\,\,\,\,,
&|{\theta_\bR}|<{\pi\over2}\,,
\\ \\
%{4\sqrt{{\lambda |z|\over\pi}}\over q_0^2\Upsilon_x}\Upsilon\equiv
\Upsilon_2(\pi-\theta_\bR)\,\,\,\,\,\,\,\,\,,
&|{\theta_\bR}|>{\pi\over2}\,,
%\\ 
\end{array}\right.
%\nonumber\\
\label{Up2d}
\end{eqnarray}
in two dimensions, and by
\beqn
\Upsilon_3(\theta_\bR)&=&\left\{
\begin{array}{ll}
B_{3}P_{2\over5}(\cos\theta_\bR) \,\,\,\,\,\,\,\,\,,
&{\theta_\bR}<{\pi\over2}\,,
\\ \\
%{4\sqrt{{\lambda |z|\over\pi}}\over q_0^2\Upsilon_x}\Upsilon\equiv
\Upsilon_3(\pi-\theta_\bR)\,\,\,\,\,\,\,\,\,,
&{\theta_\bR}>{\pi\over2}\,,
%\\ 
\end{array}\right.
%\nonumber\\
\label{Up3d}
\end{eqnarray}
in three dimensions, where $P_\nu$ is the generalized Legendre function of non-integer index.  Here $B_{2,3}$ are non-universal (i.e., system-dependent) constants.

This correlation function can be measured directly in a box counting experiment by correlating the number fluctuations in one box  with those in a {\it different} box separated from
the first by a displacement  
${\bf \delta r}$  whose magnitude $|{\bf \delta r}|$ is much greater than the largest linear
extent of either box. 

In the next section, I'll derive these results.

\section{Derivation}

My starting point is  the continuum theory for
a collection of self-propelled  active particles moving without momentum conservation (i.e., a \dry) introduced in refs.\ \cite{TT1, TT2, TT3, TT4, NL}. This theory takes the form of the following equations of motion for the velocity field $\bv$ and number density $\rho$  of the active 
particles:
\begin{eqnarray}
\partial_{t}
\bv+\lambda_1(\bv\cdot{\bf\nabla})\bv+
\lambda_2({\bf\nabla}\cdot\bv)\bv
+\lambda_3{\bf\nabla}(|\bv|^2) &=& U(|\bv|, \rho)\bv -{\bf\nabla} P +D_{B} {\bf\nabla}
({\bf\nabla}
\cdot \bv)% \nonumber \\&&
\nonumber \\
&&+ D_{T}\nabla^{2}\bv +
D_{2}(\bv\cdot{\bf\nabla})^{2}\bv+\bff
\label{EOM}
\end{eqnarray}
\begin{eqnarray}
{\partial\rho \over \partial
t}+\nabla\cdot({\bf v}\rho)=0
\label{conservation}
\end{eqnarray}
where all of the parameters $\lambda_i (i = 1 \to 4)$,
$U(\rho, |{\bf v}|)$, $D_{B,T,2}(\rho, |{\bf v}|)$ and the  ``isotropic Pressure'' $P(\rho,
|{\bf v}|)$ and the  ``anisotropic Pressure''$P_2 (\rho, |{\bf v}|)$
are, in general, functions of the density $\rho$ and the magnitude
$|{\bf v}|$ of the local velocity.
Since I am interested in an ordered, moving state with a non-zero average velocity,
I assume the $U$ term makes the local
$\bv$ have a nonzero magnitude $v_0$
in the steady state, by the simple expedient of
having $U>0$ for $|\bv|\equiv v<v_0$,
$U=0$ for $v=v_0$, and $U<0$ for $v>v_0$.

The diffusion constants  (or
viscosities) $D_{B,T,2}$  reflect the tendency of a localized fluctuation in the
velocities to spread out because of the coupling between
neighboring ``birds".   

  The ${\bf f}$ term is a random
driving force representing the noise. I assume it is Gaussian with
white noise correlations:
\begin{eqnarray}
   <f_{i}({\bf r},t)f_{j}({\bf r'},t')>=\Delta
\delta_{ij}\delta^{d}({\bf r}-{\bf r'})\delta(t-t')
\label{white noise}
\end{eqnarray}
   where $\Delta$ is a constant, and $i$ , $j$ denote
Cartesian components. The pressure $P$ tends, as in an equilibrium fluid, 
to maintain the local number density
$\rho({\bf r})$ at its mean value $\rho_0$,
and $\delta \rho = \rho -
\rho_0$.   The ``anisotropic pressure'' $P_2(\rho, |{\bf v}|)$ in
(\ref{EOM}) is only allowed due to the non-equilibrium nature of the
flock; in an equilibrium fluid such a term is forbidden, since Pascal's
Law ensures that pressure is isotropic. In the non-equilibrium steady
state of a flock, no such constraint applies. 

The final
equation (\ref{conservation}) is just conservation of bird number: we
don't allow our birds to reproduce or   die on the wing.
The interesting and novel results that arise when this constraint
is relaxed by allowing birth and death while the flock is moving has been 
discussed elsewhere\ \cite{Malthus}.

Since I am interested in an ordered, moving state with a non-zero average velocity,
I assume the $U$ term makes the local
$\bv$ have a nonzero magnitude $v_0$
in the steady state, by the simple expedient of
having $U>0$ for $v<v_0$,
$U=0$ for $v=v_0$, and $U<0$ for $v>v_0$.

The hydrodynamic model embodied in equations (\ref{EOM}), (\ref{conservation}),  and (\ref{white noise})  is equally valid  in both the
``disordered'' (i.e., non-moving)  and 
``ferromagnetically ordered'' (i.e.,  moving)  state. Here
I am interested in the ``ferromagnetically ordered'',
broken-symmetry phase which occurs when $U>0$ for $v<v_0$,
$U=0$ for $v=v_0$, and $U<0$ for $v>v_0$,as discussed earlier. In this state, I can expand the equations of motion  (\ref{EOM}) and (\ref{conservation}) for small fluctuations ${\bf \delta v}$ and $\delta \rho$ of the velocity and density about their mean values. That is, I write  the velocity and density fields as: 
\begin{eqnarray}
   \bv= (v_0+\delta v_
\parallel)\hat{{\bf x}}_\parallel+\bv_{\perp}~~,
\label{v fluc}
\end{eqnarray}
and 
\begin{eqnarray}
   \rho= \rho_0+\delta \rho~~,
\label{rho fluc}
\end{eqnarray}
   where
 $v_0\hat{{\bf x}}_\parallel=<\bv>$ is the spontaneous average
value of
$\bv$ in the ordered phase, and the fluctuations $\delta v_\parallel$ and ${\bf  v}_\perp$ of $\bv$ about this mean velocity along and perpendicular to the direction of the mean velocity are assumed to be small, as are the fluctuations $\delta\rho$ of the density. 

Expanding the equation of motion (\ref{EOM}) in these small quantities $\delta v_\parallel$, ${\bf  v}_\perp$and $\delta\rho$, and then eliminating the ``fast" variable $\delta v_\parallel$ proves to be quite subtle; see \cite{NL} for details. The result is a
pair of coupled equations of motion for the fluctuation ${\bf v}_{\perp}(\br,t)$ of the local velocity of the flock perpendicular to the direction of mean flock motion (which mean direction will hereafter denoted as "$\parallel$"), and the departure $\delta\rho(\br,t)$ of the density from its mean value $\rho_0$

\begin{widetext}
\begin{eqnarray}
&\partial_{t} {\bf v}_{\perp} + \gamma\partial_{\parallel} 
{\bf v}_{\perp} + \lambda \left({\bf v}_{\perp} \cdot
{\bf \nabla}_{\perp}\right) {\bf v}_{\perp} =-g_1\delta\rho\partial_{\parallel} 
{\bf v}_{\perp}-g_2{\bf v}_{\perp}\partial_{\parallel}
\delta\rho-g_3{\bf v}_{\perp}\partial_t
\delta\rho -{c_0^2\over\rho_0}{\bf \nabla}_{\perp}
\delta\rho -g_4{\bf \nabla}_{\perp}(\delta \rho^2)\nonumber\\&+
D_B{\bf \nabla}_\perp\left({\bf \nabla}_\perp\cdot{\bf v}_\perp\right)+
D_T\nabla^{2}_{\perp}{\bf v}_{\perp} +
D_{\parallel}\partial^{2}_{\parallel}{\bf v}_{\perp}+\nu_t\partial_t{\bf \nabla}_{\perp}\delta\rho+\nu_\parallel\partial_\parallel{\bf \nabla}_{\perp}\delta\rho+{\bf f}_{\perp} ~, 
\label{vEOMbroken}\\
%\end{eqnarray}
%\end{widetext}
%\begin{widetext}
%\begin{eqnarray}
&\partial_t\delta
\rho +\rho_o{\bf \nabla}_\perp\cdot{\bf v}_\perp
+\lambda_{\rho}{\bf \nabla}_\perp\cdot({\bf v}_\perp\delta\rho)+v_2
\partial_{\parallel}\delta
\rho =D_{\rho\parallel}\partial^2_\parallel\delta\rho+D_{\rho v} \partial_{\parallel}
\left({\bf \nabla}_\perp \cdot {\bf v}_{\perp}\right)+\phi\partial_t\partial_\parallel\delta\rho
+w_1 \partial_\parallel(\delta\rho^2)\nonumber\\&+w_2\partial_\parallel(|
{\bf v}_\perp|^2)~, \nonumber \\
\label{cons broken}
\end{eqnarray}
\end{widetext}
where  $\gamma$, $\lambda$,  $\lambda_\rho$,
$c_0^2$, $g_{1,2,3,4}$, $w_{1,2}$,  $D_{B,T,\parallel, \rho\parallel, \rho v}$, $\nu_{t,\parallel}$, $v_2$, $\phi$,  and $\rho_0$  are all phenomenological constants, 
which can be expressed in terms of the expansion coefficients of the various functions of $|\bv|$ and $\rho$ in (\ref{EOM}). The interested reader is referred to \cite{NL} for those expressions.

The important fact about these equations is that they have many non-linearities that are relevant in the renormalization group (RG) sense.
What ``relevant in the RG sense'' means in plain English is that these
specific  non-linear terms lead to  different scaling behavior at long
distances and times than predicted by the linearized version of those equations, which of course drop those
terms. This modified scaling can be encapsulated by three anomalous
scaling exponents: a ``dynamical'' exponent $z$, an ``anisotropy''
exponent $\zeta$, and a ``roughness'' exponent $\chi$ which
characterizing the scalings of time $t$, $r_\parallel$ (distance along the
direction of flock motion), and velocity ${\bf v}_\perp$
with distance $r_\perp$ perpendicular to the direction of flock motion.

It proves prohibitively difficult to calculate these universal exponents 
$z$, $\zeta$, and $\chi$ in $d < 4$, where the non-linear effects
described above become important.  
However, if one is willing to {\it conjecture} that the dominant 
non-linearity in Eqn.\ (\ref{vEOMbroken}) is $\lambda ^0_1$, and that {\it
all} of the other non-linearities are {\it irrelevant}, in the RG sense,
below
$d = 4$ (a possibility which is  not ruled out by any
calculation that has been done), then one can show
\cite{NL} that the exponents $z$, $\zeta$, and $\chi$ are
given {\it exactly} , in $d=2$, by
\begin{eqnarray}
z = {6 \over 5}
\label{zcan}
\end{eqnarray}
\begin{eqnarray}
\zeta =  {3\over 5}
\label{zetcan}
\end{eqnarray}
\begin{eqnarray}
\chi = -{1\over 5} \quad .
\label{chican}
\end{eqnarray}

Furthermore, it can also be shown that for both ``Malthusian" flocks\cite{Malthus}
(that is, \dry s  without number conservation due to `` birth and death" of the active particles) and incompressible \dry s\cite{isoincomp} , for 
all dimensions $2 \leq d \leq 4$, these exponents are given by
\begin{eqnarray}
z = 2{(d + 1) \over 5}
\label{zcan}
\end{eqnarray}
\begin{eqnarray}
\zeta =  {d + 1 \over 5}
\label{zetcan}
\end{eqnarray}
\begin{eqnarray}
\chi =  {3 -2d  \over 5} \quad .
\label{chican}
\end{eqnarray}
Note that these reduce to the values obtained by the aforementioned conjecture in $d=2$ for number conserving \dry s. 

It is therefore tempting to  conjecture that these ``canonical''  exponents (\ref{zcan}),
(\ref{zetcan}), and (\ref{chican}) apply for compressible, number conserving flocks as well.  For the remainder of this paper, I will do so, and use their values to obtain the
scaling of real space and time density and number fluctuations for $2
\leq d \leq 4$.  

In general, the equal-time,
spatially Fourier-transformed density-density correlation function predicted by these equations is\cite{NL}
\begin{eqnarray}
C_{\rho} ({\bf q}) \equiv \left<\left| \rho \left({\bf q},t
\right)\right|^2 \right>
\label{CrhoqET1}
\end{eqnarray}
is given by :
\begin{eqnarray}
C_{\rho} ({\bf q}) = 
%\int^{\infty}_{-\infty} {d\omega \over 2 \pi}
%C_{\rho} ({\bf q},\omega ) = 
{q^{2(1-\zeta)} _{\perp}  \over (\gamma-v_2)^2q_\parallel^2 +c^2q_\perp^2
}
f\left(\frac{q_\parallel}{\Lambda^{1-\zeta}q_\perp^\zeta}\right)&,
\label{CrhoqET2}
\end{eqnarray}
where the scaling 
function $f(x)$ has the limits\cite{TT4,NL}
%\begin{eqnarray}
%q_{\parallel} \gg \Lambda \left( {q_{\perp}\over 
%\Lambda}\right)^{\zeta} \quad ,
%\label{cross}
%\end{eqnarray}
\begin{eqnarray}
f(x)\to \left\{
\begin{array}{ccc}
\mathrm{constant}  &, &x \ll 1      \\
\mathrm{a}\, \mathrm{different}\, \mathrm{constant}\times x^2  &, &\,\,\,\,\,\,\,x \gg 1 \quad .\\
\end{array}
\right.
\label{scalef}
\end{eqnarray}
where $\Lambda$ is an ultraviolet
cutoff wavevector\cite{foot0}.  This scaling function is proportional to the noise strength $\Delta$; see \cite{NL} for more details.

Multiplying both sides of  (\ref{CrhoqET2}) by $(\gamma-v_2)^2q_\parallel^2 +c^2q_\perp^2$ and Fourier transforming back to real space shows that $C_\rho(\br)$ obeys an anisotropic Poisson equation
\begin{eqnarray}
\left[(\gamma-v_2)^2\pp_\parallel^2 +c^2\nabla_\perp^2\right]C_{\rho} ({\bf r}) = \nabla_\perp^2G ({\bf r})  \,,
\label{APoisson}
\end{eqnarray}
where the source term $G(\br)$ is the Fourier transform of $q^{-2\zeta(d)} _{\perp}  f\left(\frac{q_\parallel}{\Lambda^{1-\zeta}q_\perp^\zeta}\right)$. That is,
\begin{eqnarray}
G ({\bf r}) = \int {d^{d-1}q_\perp dq_\parallel \over (2\pi)^d} q^{-2\zeta(d)} _{\perp}  f\left(\frac{q_\parallel}{\Lambda^{1-\zeta}q_\perp^\zeta}\right)e^{i[{\bf q}_\perp 
\cdot {\bf r}_\perp+q_\parallel r_\parallel]}  &.
\label{G1}
\end{eqnarray}
It is straightforward to show from this expression that $G(\br)=G(\brp, \rpa)$ itself has a simple scaling form. To see this, make the linear change of variables of integration in (\ref{G1}) from $\bqp$ and $\qpa$ to ${\bf Q}_\perp$ and $Q_\parallel$ defined via $\qpa\equiv {Q_\parallel\over |\rpa|}$
and $\bqp\equiv{\Lambda {\bf Q}_\perp\over (\Lambda |\rpa|)^{1/\zeta}}$. This gives
\begin{eqnarray}
G ({\bf r}) =  |\rpa|^{(1-d)/\zeta+1}h\left({\Lambda r_\perp\over (\Lambda 
|\rpa|)^{1/\zeta}}\right)=  |\rpa|^{2\chi/\zeta}h\left({\Lambda r_\perp\over (\Lambda
|\rpa|)^{1/\zeta}}\right) \,,
\label{Gscale1}
\end{eqnarray} 
where the scaling function 
\begin{eqnarray}
h(x)\equiv
\int {d^{d-1}Q_\perp dQ_\parallel \over (2\pi)^d} Q^{-2\zeta} _{\perp}  f\left(\frac{Q_\parallel}{Q_\perp^\zeta}\right)e^{i[{\bf Q}_\perp 
\cdot \hrp x+Q_\parallel ]} \Lambda^\varpi &,
\label{h1}
\end{eqnarray}
with the utterly unimportant exponent $\varpi=(1-1/\zeta)(d-1-2\zeta)$. In deriving the second equality in (\ref{Gscale1}), I've used the values of the canonical exponents to obtain $(1-d)/\zeta+1=2\chi/\zeta$, as the algebraically inclined reader can verify for herself using the expressions (\ref{zcan}), (\ref{zetcan}), and (\ref{chican}) for the canonical values of the exponents.

Note that I expect $G(\br)$ to depend only on $\rpa$ when $\Lambda|\rpa|\gg(\Lambda\rp)^\zeta$, as it is at large $\br$ for almost all directions of $\br$. Hence, for most directions of $\br$, $\nabla_\perp^2G ({\bf r})$ vanishes. The only exception to this is the thin sliver $|\rpa|\lesssim(\Lambda\rp)^\zeta\Lambda^{-1}$, which gets very thin compared to $\rp$ for $\rp\gg\Lambda^{-1}$, since $\zeta<1$ for all $d<4$. 

This means that  
our anisotropic Poisson equation (\ref{APoisson}) has as a source on the right hand side a thin layer of charge lying very near the plane (or  the line, in $d=2$) perpendicular to the mean velocity. I will therefore model it as an {\it infinitesimally} thin layer of charge, with charge density given by
\beqn
\sigma(r_\perp)=\int_{-\infty}^\infty \nabla_\perp^2G(\rp, \rpa) d\rpa\,.
\label{sigmadef}
\eeqn
Using the scaling expression (\ref{Gscale1}) for $G(\br)$, I find
\beqn
\nabla_\perp^2G(\rp, \rpa)&=& |\rpa|^{2\chi/\zeta}\nabla_\perp^2h\left({\Lambda r_\perp\over (\Lambda|\rpa|)^{1/\zeta}}\right)=|\rpa|^{2\chi/\zeta}{\Lambda^2\over(\Lambda|\rpa|)^{2/\zeta}}\left[{(d-2)\over u}h'\left(u\right)+h''(u)\right]\nonumber\\
&\equiv&|\rpa|^{-4}Y(u) \,,
\label{Lap G}
\eeqn
where I've defined the scaling variable
\beq
u\equiv{\Lambda r_\perp\over (\Lambda|\rpa|)^{1/\zeta}}
\label{udef}
\eeq
and the scaling function
\beq
Y(u)\equiv\Lambda^{2(1-1/\zeta)}\left[{(d-2)\over u}h'\left(u\right)+h''(u)\right] \,.
\label{Ydef}
\eeq
I've also 
used the fact that ${2(\chi-1)\over\zeta}=-4$, as the skeptical reader can verify for himself by once again using the expressions (\ref{zcan}), (\ref{zetcan}), and (\ref{chican}) for the canonical values of the exponents.

Using the last equality in (\ref{Lap G}) in my expression (\ref{sigmadef}) for $\sigma(r_\perp))$ gives
\beqn
\sigma(r_\perp)=2\int_0^\infty Y(u)\rpa^{-4} d\rpa\,,
\label{sigma2}
\eeqn
where the scaling variable $u$ continues to be given by (\ref{udef}), and I've used the fact that the scaling function $Y(u)$ is an even function of $\rpa$ to replace the integral over $\rpa$ over the range  [$-\infty\,,\infty$] with twice the integral over the range [$0\,,\infty$].

Solving (\ref{udef}) for $\rpa$ in terms of $u$ gives (for positive $\rpa$, which is all I need) 
\beq
\rpa=\Lambda^{-1}\left({\Lambda\rp\over u}\right)^\zeta \,.
\label{rpa(a)}
\eeq
Now changing variables of integration in (\ref{sigma2}) from $\rpa$ to $u$ (keeping in mind that the integral over $\rpa$ is at constant $\rp$) gives
\beqn
\sigma(r_\perp)=A\rp^{-3\zeta}\,,
\label{sigmascale}
\eeqn
where I've defined the constant
\beqn
A=2\int_0^\infty Y(u)u^{3\zeta-1} du\,.
\label{Adef}
\eeqn

Note that this constant is non-universal - that is, it depends on the hydrodynamic parameters of the  particular flock we're studying (through both the noise strength $\Delta$ and the ultraviolet cutoff $\Lambda$). But for a {\it given type of flocker}, it is independent of position$\br$, time $t$, and the size of the flock.

Thus, our anisotropic Poisson equation (\ref{APoisson})  can be rewritten as 
\begin{eqnarray}
\left[(\gamma-v_2)^2\pp_\parallel^2 +c^2\nabla_\perp^2\right]C_{\rho} ({\bf r}) = A\rp^{-3\zeta}\delta(\rpa)  \,.
\label{Asheet}
\end{eqnarray}
I recognize that at this point the more skeptical reader may be doubting this ``infinitely thin sheet" approximation. In particular, she may be wondering whether the value of $C_\rho(\br)$ found this way might be invalid within the region 
\beq
|\rpa|\lesssim(\Lambda\rp)^\zeta\Lambda^{-1} \,,
\label{sheet}
\eeq
within which the ``charge" distribution does not look like a thin sheet. However, I will show later that this is no more a problem here than it is for real electrostatic problems involving charge layers, for which it is not necessary to consider the finite thickness of a real charge layer, since the ``potential" cannot change appreciably over that thickness. I will likewise show a posteriori here that the correlation function $C_\rho(\br)$ does not change appreciably (for large $\br$) over the thin sheet
(\ref{sheet}); hence, I can use the result of this thin sheet
 calculation for {\it all} $\br$, even those within the sheet.
 
 A consequence of this, as we'll see, is that even though $C_\rho(\bq)$ is, as equation (\ref{CrhoqET2}) shows, strongly anisotropic- indeed, it exhibits anisotropic scaling- $C_\rho(\br)$ is nearly isotropic, and in particular is completely isotropic in its scaling.
 
 Before proceeding, in the interests of making the analogy with electrostatics more perfect, I will anisotropically rescale lengths to make (\ref{Asheet}) an isotropic Poisson equation. Specifically, I'll define a new vector $\bR$ via
 \beqn
 \bRp=\brp\,,\,\Rpa={c\rpa\over
 |\gamma-v_2|}
 \label{R rescale}
 \eeqn
 In this new variable $\bR$, equation (\ref{Asheet})
becomes a completely isotropic Poisson equation:
\begin{eqnarray}
\nabla_\bR^2C_{\rho} ({\bf R}) = A'\Rp^{-3\zeta}\delta(\Rpa)  \,,
\label{Isheet}
\end{eqnarray}
where I've defined $A'={A\over |\gamma-v_2|c}$.

By inversion symmetry,  $C_\rho(\bR)$ must remain unchanged when $\Rpa\to-\Rpa$.  This will lead to a gradient discontinuity in $C_\rho(\bR)$ at the equatorial plane $\theta=\pi/2$. By the usual ``Gaussian pillbox" argument of  electrostatics, the presence
of a charge sheet is equivalent to a boundary condition at the equatorial plane:
\beqn
(\nabla_\bR)_N C_\rho(\bR)=-{1\over R}\left({\partial C_\rho\over\partial\theta_\bR}\right)^-(R,\theta_\bR=\pi/2)={A'\over2}R^{-3\zeta} \,.
\label{BC}
\eeqn
Here the superscript ``$-$" on $\left({\partial C_\rho\over\partial\theta_\bR}\right)^-(R,\theta_\bR=\pi/2)$ denotes a derivative evaluated as $\theta_\bR\to\pi/2$ from below. The derivative as $\theta_\bR\to\pi/2$ from above has the opposite sign, due to the inversion symmetry of $C_\rho$.

So now I must satisfy the Poisson equation (\ref{Isheet}) subject to the boundary condition (\ref{BC}). I will seek a separable solution of the form
\beqn
C_\rho(\bR)=R^{-\alpha}\Upsilon_d(\theta_\bR) \,.
\label{trial}
\eeqn

By the inversion symmetry of  $C_\rho(\bR)$, I know that 
\beq
\Upsilon_d(\pi-\theta_\bR)=\Upsilon_d(\theta_\bR) \,.
\label{Upsidedown}
\eeq
This will lead to a slope discontinuity in $\Upsilon_d(\theta_\bR)$ at $\theta=\pi/2$, which is, of course, precisely what is generated by the thin charge layer.
 
Inserting the ansatz (\ref{trial}) into the boundary condition (\ref{BC}) giuves
\beqn
R^{-\alpha-1}\Upsilon_d'(\theta_\bR=\pi/2)={A'\over2}R^{-3\zeta} \,,
\label{BCtrial}
\eeqn
which implies
\beq
\alpha+1=3\zeta \,.
\label{alpha1}
\eeq
This is obviously trivially solved to give:
\beq
\alpha(d)=3\zeta(d)-1={3d-2\over5} \,,
\label{alpha}
\eeq
where in the last equality I have used the canonical value (\ref{zetcan}) for $\zeta(d)$.

So far, I have worked in completely general spatial dimension $d$. To proceed, I'll now deal specifically with the two physical cases $d=2$ and $d=3$.

In $d=2$,  (\ref{alpha}) gives $\alpha={4\over5}$. Requiring that the ansatz (\ref{trial}) obeys Laplace's equation (\ref{Isheet}) away from the plane $\Rpa=0$ determines $\Upsilon_2$:
\beqn
\Upsilon_2(\theta_\bR)=B_{2}\cos(\alpha\theta_\bR)=B_{2}\cos(4\theta_\bR/5) \,,
\label{Up2d2}
\eeqn
which is the result (\ref{Up2d}) quoted in the introduction.
Fixing $B_{2}$ using the boundary condition (\ref{BC}) gives
\beqn
B_{2}={A'\over\sin({2\pi\over5})}\approx1.05A' \,.
\label{B2d}
\eeqn

Note that $C_\rho(\bR)$ is identical in the upper half plane (i.e., $-{\pi\over2}<\theta_\bR<{\pi\over2}$) to the solution for the electrostatic potential near a sharp upward pointing conducting wedge\cite{Jackson} of opening angle ${3\pi\over8}=67.5^\circ$. Of course, in the lower half plane $C_\rho(\bR)$ is just the mirror image of the upper half plane solution, since  $C_\rho(\bR)$ is symmetric about the axis $\Rpa=0$. 

In $d=3$, I have $\alpha={7\over5}$, and requiring that the ansatz (\ref{trial}) obeys Laplace's equation away from the plane $\Rpa=0$ determines $\Upsilon_3$:
\beqn
\Upsilon_3(\theta_\bR)=B_{3}P_{\alpha-1}(\cos\theta_\bR)=B_{3}P_{2\over5}(\cos\theta_\bR) \,,
\label{Up3d2}
\eeqn
where $P_\nu$ is the generalized Legendre function of non-integer index. This is, of course, just the result (\ref{Up3d}) quoted in the introduction.

Fixing $B_{3}$ using the boundary condition (\ref{BC}) gives
\beqn
B_{3}=-{A'\over \left({dP_{2\over5}(\theta)\over d\theta}\right)_{\theta={\pi\over2}}}=-{5\Gamma\left({17\over10}\right)\Gamma\left(-{1\over5}\right)\over7\sqrt{\pi}}A'\approx2.13156A' \,.
\label{B3d}
\eeqn

Note that $C_\rho(\bR)$ is now identical in the upper half space (i.e., $0<\theta_\bR<{\pi\over2}$) to  the electrostatic potential near a sharp upward pointing charged cone\cite{Jackson} of opening angle $37.16^\circ$. Of course, in the lower half space  $C_\rho(\bR)$ is just the mirror image of the upper half space solution, since $C_\rho(\bR)$ is symmetric about the plane $\Rpa=0$.

Using the coordinate transformation (\ref{R rescale}) to rewrite the above results in terms of the real coordinates $\br$, I have 
\begin{eqnarray}
 C_{\rho} ({\bf r}) = r^{-\alpha(d)} G_d(\theta_{{\bf r} })
\label{Crhorscale}
\end{eqnarray}
where $r$ is the magnitude of  ${\bf r}$ $ ( r=|{\bf r}| )$, 
$\theta_{{\bf r}}$ is the angle between  ${\bf r}$ and the direction of
mean flock motion
$\hat{{\bf x}}_\parallel$,  
\begin{eqnarray}
G_d(\theta_{{\bf r}}) \equiv {\Upsilon_d(\theta_{{\bf R}})\over\left[{c^2\over (\gamma-v_2)^2}\cos^2\theta+\sin^2\theta\right]^{\alpha(d)/2}} \,,
\label{Gdef2}
\end{eqnarray}
which is just equation (\ref{Gdef}) of the introduction, with
\beqn
\theta_\bR=\tan^{-1}\left({\Rp\over\Rpa}\right)=\tan^{-1}\left({|\gamma-v_2|\over c}{\rp\over\rpa}\right)=\tan^{-1}\left({|\gamma-v_2|\over c}\tan(\theta_\br)\right) \,,
\label{thetaR2}
\eeqn
which is just (\ref{thetaR}) of the introduction, with
the function $\Upsilon_d$ given by equations (\ref{Up2d}) and (\ref{Up3d}) quoted in the introduction for $d=2$ and $d=3$ respectively, 
and $\alpha(d)$ given by (\ref{alpha}).

This result summarized by (\ref{Crhorscale}), (\ref{alpha}), (\ref{Up2d}), (\ref{Up3d}), and (\ref{Gdef}) for the real space, equal-time density-density correlation function  are the basis of derivation of the giant number fluctuations I am about to perform. To complete that  derivation, I must first complete the {\it a posteriori} argument made earlier that the departures of $C_\rho$ from the ``infinitely thin sheet" approximation are negligible. 

This is quite straightforward to do using the electrostatic analogy. Within the sheet, whose thickness $|\rpa|$, I remind the reader, is given by (\ref{sheet}), which says $
|\rpa|\lesssim(\Lambda\rp)^\zeta\Lambda^{-1}$, the ``electric field" $(\nabla_\bR)_N C_\rho(\bR)$ will always 
be less, by the ``Gaussian pillbox" argument, than that just above the thin sheet, since a Gaussian pillbox that starts at the equatorial plane and ends within the thin sheet will always contain {\it less} charge than on that spans the entire thickness of the sheet. Therefore, the ``potential" - which is actually the correlation function $C_\rho$ - can change within the sheet by no more than $(\nabla_\bR)_N C_\rho(\bR)$ evaluated just outside the sheet,  times the thickness of the sheet. Since $(\nabla_\bR)_N C_\rho(\bR)\sim C_\rho/\rp$, and the thickness of the sheet is $
|\rpa|\lesssim(\Lambda\rp)^\zeta\Lambda^{-1}\ll\rp$, the last inequality holding for all $\rp\gg\Lambda^{-1}$ since $\zeta={d+1\over5}<1$ for all $d<4$, it follows that the change $\delta C_\rho$  in the ``potential", - that is, in $C_\rho$- across the thickness of the thin sheet obeys $\delta C_\rho<((\nabla_\bR)_N C_\rho(\bR)|\rpa|\lesssim  C_\rho/\rp)|(\Lambda\rp)^\zeta\Lambda^{-1}\propto C_\rho\rp^{\zeta-1}$. Since $\zeta<1$, this is much less than $C_\rho$ itself, so the change in $C_\rho$ across the thickness of the thin sheet is indeed negligible, as I assumed.

Since $\alpha(d=2)={4\over5}=.8$ ,
and
$\alpha(d=3)={7\over5}=1.4 $,
are quite different, it would appear to be quite straightforward to  see the difference
between the scaling behavior of density fluctuations in two and three
dimensions in simulations or experiments.

Unfortunately, things are not quite so simple.  The most natural 
quantity to look at when studying density fluctuations is the
fluctuations of the number of particles in an imaginary ``counting box'' (which need not be a rectangular, or even polyhedral, but could, for example, be a (hyper)sphere or an ellipsoid, etc.) of some
volume $V_{\rm box}$ inside a flock of volume $V_{\rm flock} \gg V_{\rm
box}$.  The mean squared number fluctuations $\left<(\delta N)^2 \right>
\equiv \left< N^2\right> - \left< N\right>^2$ can readily be related to
the real space correlations $C_{\rho} ({\bf r})$:
\begin{eqnarray}
\left<(\delta N)^2 \right> &=&\int_V d^dr d^dr^{\prime}\left< \delta  \rho
({\bf r})\delta \rho ({\bf r}^{\prime})\right>\\ \nonumber &=&\int_{V}d^dr
d^dr^{\prime} \, C_{\rho}\left( {\bf r} - {\bf r}^{\prime}\right)
\label{delN1}
\end{eqnarray}
where the subscript $V$ denotes that the integrals are over ${\bf r}$ 
and ${\bf r}^{\prime}$'s contained within our experimental ``counting box''. Using
our expression (\ref{Crhorscale}) for $C_{\rho}\left( {\bf r} -
{\bf r}^{\prime}\right)$ gives
\begin{eqnarray}
\left<(\delta N)^2 \right> &=&\int_V d^dr d^dr^{\prime}   |{\bf r}- 
{\bf r}^{\prime}|^{-\alpha (d)} G_d \left(\theta_{{\bf r} -
{\bf r}^{\prime}} \right)
\label{delN2}
\end{eqnarray}
Now let's take our ``box'' to be an arbitrary shape with total volume $V=L^d$. 
%$d$-dimensional hypercube of  side $L$ (e.g., on $L \times L$ square in $d = 2$, or an $L \times L \times L$ cube in $d = 3$) with one of its faces perpendicular to the direction of mean flock motion.  
Making the changes of variables ${\bf r} \equiv
{\bf x}L, {\bf r}^{\prime} \equiv {\bf x}^{\prime}L$, I obtain
\begin{eqnarray}
\left<(\delta N)^2 \right> = L^{2d - \alpha (d)} \int_{V_1}d^dx 
d^dx^{\prime}  |{\bf x} - {\bf x}^{\prime}\: |^{ - \alpha (d)}  G_d
\left(\theta_{{\bf x} - {\bf x}^{\prime}} \right)
\label{delN3}
\end{eqnarray}
where $V_1$ denotes that the integrals are over ${\bf x}$ and 
${\bf x}^{\prime}$ contained in a {\it unit} volume of the same shape as our original counting box.
%hypercube.  Hence
Clearly, this
integral has {\it no} dependence on $L$.  Therefore (\ref{delN3}) implies
\begin{eqnarray}
\left<(\delta N)^2 \right> = L^{2d - \alpha (d)}  \times K\left({c\over |\gamma-v_2|},\rm{shape}\right)
\label{delN4}
\end{eqnarray}
where the constant 
\beqn
K\left({c\over |\gamma-v_2|},\rm{shape}\right)
\equiv \int_{V_1}d^dx 
d^dx^{\prime}  |{\bf x} - {\bf x}^{\prime}\: |^{ - \alpha (d)}  G_d
\left(\theta_{{\bf x} - {\bf x}^{\prime}} \right)
\label{Kshape}
\end{eqnarray}
depends on the {\it shape} of the box (as well as the ratio ${c\over |\gamma-v_2|}$, which enters both explicitly in equation (\ref{Gdef}) for $G_d$ and implicitly through the relation (\ref{thetaR})  between $\theta_\br$ and $\theta_\bR$),
 but is independent of its size $L$.  This can be rewritten in 
terms of the mean number $\left<N\right>$ of critters in the counting box, using
the fact that the average density $\rho _0$ is well-defined. Hence,
$\left<N\right> = \rho _0 L^d$, or $L = \left({\left<N\right> \over\rho
_0 } \right)^{1 \over d}$.  Using this in (\ref{delN4}) and taking the
square root of both sides gives:
\begin{eqnarray}
\sqrt {<(\delta N)^2>} = K'<N>^{\phi(d)}
\label{Nfluc2}
\end{eqnarray}
with
\begin{eqnarray}
\phi(d)=\frac{2d-\alpha(d)}{2d}=\frac{7}{10}+\frac{1}{5d}  \,.
\label{phi}
\end{eqnarray}
The coefficient
\beq
K'\equiv\rho_0^{-\phi(d)}\sqrt{K\left({c\over |\gamma-v_2|},\rm{shape}\right)} 
\label{K'}
\eeq
also depends on the shape of the box and the ratio ${c\over |\gamma-v_2|}$. Equation (\ref{Nfluc2}) is just (\ref{Nfluc}) of the introduction.

Eqn.(\ref{phi}) gives
\begin{eqnarray}
\phi(d=2) = .8
\label{phi2}
\end{eqnarray}
and
\begin{eqnarray}
\phi(d=3) = 23/30 = .7666666......
\label{phi3}
\end{eqnarray}

Note that in {\it all} dimensions $d$, even $d>4$, where there is no
``anomalous hydrodynamics'', the scaling of number  fluctuations with mean
number violates the ``law of large numbers'': the general rule that rms
number fluctuations scale like the square root of mean number.   The
fluctuations eqn.
(\ref{Nfluc}) are infinitely larger than this prediction in the limit of
mean number 
$\left<N\right> \to \infty$ for {\it all} spatial dimensions $d$; hence,
they are much larger than those found in most equilibrium \cite{BEC}
and most non-equilibrium systems, since most of those obey the law of
large numbers.  Giant number fluctuations like those found here, but even
larger, are predicted theoretically
\cite{actnemsub} and observed experimentally \cite{Nemexp} in ``nematic''
flocks, in which active creature align their long axes, but are equally
likely to be moving in either direction along that axis, so that the net
velocity is zero.

In addition to obeying a different scaling law, 
number fluctuations in \dry s  exhibit another phenomenon not present in most other systems: the number fluctuations depend not only on the mean number $<N>$  of particles in the box, but also on its shape, as embodied in the coefficient $K'=\sqrt{K\left({c\over |\gamma-v_2|},\rm{shape}\right)}$ in (\ref{Nfluc}). %To see this, consider counting particles in a rectangular box in $d=2$, or a rectangular parallelepiped box in $d=3$ (or, for that matter, a {\it hyper}parallelepiped box in $d>3$) whose length $L_\parallel$ along the direction of mean flock motion is different from its length $L_\perp$  in the other $d-1$ directions. Let's take $L_\perp=\beta L$ and $L_\parallel=L/\beta^{d-1}$. I've chosen these dimensions so that the (hyper-)volume of this box is the same as the volume of the (hyper-)cubic box I considered earlier (i.e., $L^d$). Hence, the mean number of particles $<N>=\rho_0L^d$ in this box is the same as that in the earlier cubic box.

%But the mean squared number fluctuations are not. To see this, simply proceed as I just did, making the changes of variables ${\bf r} \equiv {\bf x}L, {\bf r}^{\prime} \equiv {\bf x}^{\prime}L$, to obtain \begin{eqnarray}\left<(\delta N)^2 \right> = L^{2d - \alpha (d)} \int_{V_2}d^dx d^dx^{\prime}  |{\bf x} - {\bf x}^{\prime}\: |^{ - \alpha (d)}  G_d\left(\theta_{{\bf x} - {\bf x}^{\prime}} \right)\label{delN3}\end{eqnarray}
%where $V_2$ denotes that the integrals are over ${\bf x}$ and ${\bf x}^{\prime}$ contained in a hyper-parallelepiped of unit {\it volume}, but a different shape than the simple unit cubic I obtained earlier. Specifically,the integral over each Cartesian component $R^\perp_i$ of ${\bf x}_\perp$ runs  over the interval $0<R^\perp_i<\beta$, while the integral over $R_\parallel$ runs over $0<R_\parallel<\beta^{1-d}$. The same limits apply to the integral over ${\bf x}'$, of course. 

%  Hence, while this integral has  no dependence on {\it $L$}, it {\it does} depend on $\beta$ (since the integrand is not isotropic). Hence, the coefficient of   $ L^{2d - \alpha (d)}$ in (\ref{delN3}) , and, as a result, the coefficient of $<N>^{\phi(d)}$ in (\ref{Nfluc}), will depend on $\beta$, or, equivalently, the aspect ratio of the box. More generally, the coefficient will depend upon the shape of the box. 

This dependence is singular in the limit of  a ``needle shaped" counting box;  that is, one that is much longer along the direction $\hat{{\bf x}}_\parallel$ of flock motion than perpendicular to it. I mean singular in the sense that the coefficient $K'=K\left({c\over |\gamma-v_2|},\rm{shape}\right)$ actually {\it vanishes} in the limit that the aspect ratio $\beta\equiv{L_\parallel\over L_\perp}$ of the box goes to infinity ($\beta\to\infty$), where $L_\parallel$ and $L_\perp$ are respectively the linear extents of the counting box along and perpendicular to the direction of flock motion. I will illustrate this first in $d=3$, with the example of a  counting box that is a cylinder with its axis along  the $\hat{{\bf x}}_\parallel$ direction, with height $L_\parallel$ and radius $L_\perp$. The volume of this cylinder is clearly $\pi L_\perp^2L_\parallel$, and, hence, the mean number of particles in it is
\beqn
<N>_{\rm cylinder}=\rho_0\pi L_\perp^2L_\parallel =\rho_0\pi L_\perp^3 \beta\,.
\label{ncyl}
\eeqn%since the anisotropic  function $G_d (\theta)$ is finite and non-zero in all directions, it should be observable in experiments and simulations.

Our general expression expression (\ref{delN2}) for $\left<(\delta N)^2 \right>$ reads for this case
\begin{eqnarray}
\left<(\delta N)^2 \right> &=&\int_{\rp<L_\perp} d^2\rp \int_{\rp'<L_\perp}d^2\rp^{\prime}  \int_0^{L_\parallel} d\rpa  \int_0^{L_\parallel} d\rpa' \,\,|{\bf r}- 
{\bf r}^{\prime}|^{-7/5} G_3 \left(\theta_{{\bf r} -
{\bf r}^{\prime}} \right) \,.
\label{delNcyl1}
\end{eqnarray}

I note that the integrals over $\rpa$ and $\rpa'$ both {\it converge} in the limit $L_\parallel\to\infty$. This follows from the fact that the integrand $|{\bf r}- 
{\bf r}^{\prime}|^{-7/5} G_3 \left(\theta_{{\bf r} -
{\bf r}^{\prime}} \right)\propto
\rpa^{-7/5}$ as $\rpa\to\infty$, and likewise for $\rpa'$. This result also uses the fact that $G_3(\theta)$ is finite and  non-zero for all $\theta$, and in particular for $\theta\to0$.

Since this falloff with $\rpa$ and $\rpa'$ is faster than $1/r_\parallel$, the integrals over $\rpa$ and $\rpa'$ both converge in the limit $\L_\parallel\to\infty$. Note that this will 
{\it not} be true in $d=2$, where $\alpha(2)=4/5<1$, so the analogous integral will {\it not} converge.

This means that if the aspect ratio $\beta$ is $\gg1$ - that is, if $\L_\parallel\gg L_\perp$ - I can accurately approximate the value of the integral in (\ref{delNcyl1}) by taking $L_\parallel\to\infty$. Thus I get
\begin{eqnarray}
\left<(\delta N)^2 \right> &=&\int_{\rp<L_\perp} d^2\rp \int_{\rp'<L_\perp}d^2\rp^{\prime}  \int_0^{\infty} d\rpa  \int_0^{\infty} d\rpa' \,\,|{\bf r}- 
{\bf r}^{\prime}|^{-7/5} G_3 \left(\theta_{{\bf r} -
{\bf r}^{\prime}} \right) \,.
\label{delNcyl2}
\end{eqnarray}
I can now evaluate this integral by making a very similar change of variables to that used earlier; namely ${\bf r} \equiv
{\bf x}L_\perp, {\bf r}^{\prime} \equiv {\bf x}^{\prime}L_\perp$. I thereby obtain
\begin{eqnarray}
\left<(\delta N)^2 \right> =K_{\rm cyl}L_\perp^{23/5}
\label{delNcyl3}
\eeqn
where I've defined 
\beqn
K_{\rm cyl}\equiv\int_{|{\bf x}_\perp|<1} d^2x_\perp \int_{|{\bf x}_\perp|<1}d^2x_\perp^{\prime}  \int_0^{\infty} dx_\parallel  \int_0^{\infty} dx_\parallel ' \,\,|{\bf x}- 
{\bf x}^{\prime}|^{-7/5} G_3 \left(\theta_{{\bf x} -
{\bf x}^{\prime}} \right) \,,
\label{delNcyl4}
\end{eqnarray}
which I remind the reader is a perfectly finite function of the ratio ${c\over |\gamma-v_2|}$ (which is buried in $G_3$). It is also independent of the aspect ratio $\beta$.

Solving my expression (\ref{ncyl}) for $L_\perp(<N>, \beta)$ gives 
\beq
L_\perp=\left({<N>\over\pi\rho_0\beta}\right)^{1/3} \,.
\label{LperpN}
\eeq 
Using this, I can rewrite (\ref{delNcyl3}) (or, more precisely, its square root) in terms of $<N>$ and the aspect ratio $\beta$:
\begin{eqnarray}
\sqrt {<(\delta N)^2>} = K'<N>^{23/30}
\label{Nfluccyl}
\end{eqnarray}
with
\begin{eqnarray}
K'=\sqrt{K_{\rm cyl}}/(\pi\rho_0\beta)^{23/30}\propto\beta^{-23/30} \,,
\label{K'2}
\end{eqnarray}
which, as claimed, vanishes as the cylinder gets very long (i.e., as $\beta\to\infty$). Note also that I've recovered the general "${23\over30}$" scaling law for $\sqrt {<(\delta N)^2>}$ with $<N>$ in $d=3$.

It is straightforward to see that for most three dimensional "needle" shapes (e.g., an ellipsoid of revolution about   the $\hat{{\bf x}}_\parallel$ direction, with its long axis in that direction), the same ``$23/30$" scaling law (\ref{K'2}) for the coefficient $K'$ in (\ref{Nfluccyl}) with aspect ratio $\beta$ (which in the ellipsoid case will be the ratio of semi-major to semi-minor axis) will apply.

I can also obtain a simple expression for the  ratio the value of $K'$ for a ``pancake" shaped counting volume, by which I mean a volume much shorter
 along the direction $\hat{{\bf x}}_\parallel$ of flock motion than perpendicular to it.
 
 Consider in particular a very squat cylinder with its axis along $\hxp$. For such a shape, I can now approximate $\br$ with $\brp$, $\br'$ with $\brp'$, and $\theta_{\br-\br'}\approx{\pi\over2}$ for the range of $\br$ and $\br'$ that dominate the integral. This gives
 \begin{eqnarray}
\left<(\delta N)^2 \right> &=&G_3 \left({\pi\over2} \right) \int_0^{L_\parallel} d\rpa  \int_0^{L_\parallel} d\rpa'\int_{\rp<L_\perp} d^2\rp \int_{\rp'<L_\perp}d^2\rp^{\prime}  \,\,|{\bf r}_\perp- 
{\bf r}_\perp^{\prime}|^{-7/5}  \,.
\label{delNpan1}
\end{eqnarray}

Using the  change of variables  ${\bf r} \equiv
{\bf x}L_\perp, {\bf r}^{\prime} \equiv {\bf x}^{\prime}L_\perp$ for the $\brp$ and $\brp'$ integrals, and doing the trivial integrals over $\rpa$ and $\rpa'$,  I get
\begin{eqnarray}
\left<(\delta N)^2 \right> =K_{\rm pan}L_\perp^{13/5}L_\parallel^2
\label{delNpan2}
\eeqn
where I've defined
\beqn
K_{\rm pan}\equiv G_3 \left({\pi\over2} \right)\int_{|{\bf x}_\perp|<1} d^2x_\perp \int_{|{\bf x}_\perp|<1}d^2x_\perp^{\prime}  \,\,|{\bf x}_\perp- 
{\bf x}_\perp^{\prime}|^{-7/5}  \approx38.651A'\,,
\label{delNpan3}
\end{eqnarray}
where I've used equation (\ref{Gdef}) for $G_3$, which implies 
\beqn
G_3\left({\pi\over2}\right)=
\Upsilon_3\left({\pi\over2}\right)=B_3P_{2\over5}\left(0\right)={25\Gamma\left({17\over10}\right)\Gamma\left({4\over5}\right)\over7\Gamma\left({6\over5}\right)\Gamma\left({3\over10}\right)}A'\approx1.3755A'\,,
\eeqn
the first two 
equalities following from equations (\ref{Gdef})  and (\ref{Up3d}), respectively. The penultimate equality follows from known properties of the generalized Legendre functions\cite{GR}. I've also numerically 
evaluated the four dimensional integral displayed explicitly in (\ref{delNpan3}) (it's equal to $28.1$). 

Since I'm still dealing with a cylinder
 here, the expression (\ref{LperpN}) for for $L_\perp(N, \beta)$ continues to hold. Using this and $L_\parallel=\beta L_\perp$ in (\ref{delNpan2}) and taking the usual square root gives for the rms number fluctuations 
 \begin{eqnarray}
\left<(\delta N)^2 \right> ={\sqrt{K_{\rm pan}}\over(\pi\rho_0)^{23/30}}<N>^{23/30}\beta^{8/15} \,,
\label{delNpan4}
\eeqn
 which vanishes as $\beta\to0$. The scaling of this result with the aspect ratio $\beta$ is the result (\ref{K'pan3}) quoted in the introduction.
 
To summarize what I've shown, in three dimensions the coefficient $K'$ of $<N>^{23/30}$ in the scaling law (\ref{Nfluc}) for the rms fluctuations $\left<(\delta N)^2 \right>$ vanishes as the aspect ratio $\beta\to0$ like $\beta^{8/15}$, and as $\beta^{-23/30}$ as $\beta\to\infty$. Thus, there must be an optimal aspect ratio $\beta\sim 1$ where this coefficient is maximized. Thus, somewhat surprisingly given the anisotropy of the Fourier transformed correlation function, the optimal box for observing the largest possible giant number fluctuations proves to be roughly isotropic (e.g., a cube or a sphere). The precise value of the optimal ratio will depend on the ratio of hydrodynamic parameters ${c\over |\gamma-v_2|}$.

The same qualitative behavior with aspect ratio proves to hold in two dimensions as well. I'll show this by considering a rectangular counting box aligned with two of its edges parallel to $\hxp$. In the ``needle" limit, this will be the long axis, while for the ``pancake" limit (which in $d=2$ is just the needle rotated by $90$
 degrees), it will be the short axis. 
 
 Continuing to define $\beta\equiv{L_\parallel\over L_\perp}$ in all cases, I'll now focus first on the needle case $\beta\gg1$.

Because $\alpha=4/5<1$ in $d=2$, the double integral in (\ref{delN2}) does not converge at large $\rpa$; therefore, it that integral  is dominated by widely separated values of $\rpa$. This implies that those integrals will be dominated, for the needle geometry, by values of $\br$ and $\br'$ such that $\theta_{\br-\br'}\ll1$. Furthermore, for these values of $\br$ and $\br'$, $|\br-\br'|\approx|\rpa-\rpa'|$. With these approximations, which become exact in the limit of the aspect ratio $\beta\to\infty$, I can therefore write
\begin{eqnarray}
\left<(\delta N)^2 \right> &=&B_2\left({|\gamma-v_2|\over c}\right)^{4/5}  \int_0^{L_\parallel} d\rpa  \int_0^{L_\parallel} d\rpa'\int_0^{L_\perp} d\rp \int_0^{L_\perp}d\rp^{\prime}  \,\,|\rpa- 
\rpa^{\prime}|^{-4/5}  \,.\nonumber\\
\label{delN2dneedle1}
\end{eqnarray}
All of the integrals in this expression are elementary; doing them gives
  \begin{eqnarray}
\left<(\delta N)^2 \right> &=&JL_\perp^2L_\parallel^{6/5}  \,,%\nonumber\\
\label{delN2dneedle2}
\end{eqnarray}
where I've defined
\beq
J\equiv{25\over3}B_2\left({|\gamma-v_2|\over c}\right)^{4/5} \,.
\eeq
 To re-express the number fluctuations (\ref{delN2dneedle2}) in terms of the mean number of particles $<N>$, I can use the fact that
 \beqn
<N>_{\rm rectangle}=\rho_0 L_\perp L_\parallel =\rho_0L_\perp^2 \beta\,,
\label{nrect}
\eeqn
 which, when combined with (\ref{delN2dneedle2}) and $L_\parallel=\beta L_\perp$ gives
 \begin{eqnarray}
\sqrt{<(\delta N)^2>} = \sqrt{J}\rho_0^{-8/5}<N>^{8/5}\beta^{-1/5} \,.
\label{Nfluc2dneed}
\end{eqnarray}

For the pancake, I have
  \begin{eqnarray}
\left<(\delta N)^2 \right> &=&G_2\left({\pi\over2}\right) \int_0^{L_\parallel} d\rpa  \int_0^{L_\parallel} d\rpa'\int_0^{L_\perp} d\rp \int_0^{L_\perp}d\rp^{\prime}  \,\,|\rp- 
\rp^{\prime}|^{-4/5}=J_pL_\parallel^2L_\perp^{6/5}  \,,\nonumber\\
\label{delN2dpan1}
\end{eqnarray}
where I've defined
\beq
J_p\equiv{25\over3}G_2\left({\pi\over2}\right)={25\over6\tau}B_2 \,,
\eeq
where $\tau={\sqrt{5}+1\over2}=1.6180...$ is the Golden mean.

Using (\ref{nrect}) and $L_\parallel=\beta L_\perp$ again 
gives
 \begin{eqnarray}
\sqrt{<(\delta N)^2>} = \sqrt{J_p}\rho_0^{-8/5}<N>^{8/5}\beta^{1/5} \,.
\label{Nfluc2dpan}
\end{eqnarray}

For such a needle shaped box, almost all angles $\theta_{{\bf r} - {\bf r}^{\prime}}$between two randomly chosen points $\br$ and $\br'$ in the integrals in (\ref{delN2}) obey $\theta_{{\bf x} - {\bf x}^{\prime}}\ll1$.

Note that these giant {\it number} fluctuations are {\it not } due to giant {\it density} fluctuations. In fact, the fluctuations in the density {\it at any single point} are perfectly finite, and not necessarily bigger than those in some equilibrium systems. It is the long-ranged {\it correlations} of those density fluctuations that give rise to giant number fluctuations.

Dramatic as the large fluctuations predicted by eqn. (\ref{Nfluc}) are,
the  exponents $\phi(2)=.8$ in $d = 2$ and $\phi(3)=23/30=.766666...$ in $d = 3$ are numerically too close to each
other for the difference between the behavior in the two different
dimensions to be easily detectable experimentally.  A more direct measure of
$\alpha(d)$, which differs considerably between two ($\alpha(d=2)=.8$)
and three ($\alpha(d=3)=1.4$) dimensions, would clearly be more useful. 

One way to do so is to correlate the number fluctuations in one box  with those in a {\it different} box separated from
the first by a displacement  
${\bf \delta r}$  whose magnitude $|{\bf \delta r}|$ is much greater than the largest linear
extent of either box. 
In this case, the correlations are given by
\begin{eqnarray}
\left< \delta N_1 \delta N_2 \right> =  \int _{V_1} d^dr \int _{V_2}
d^dr^{\prime} C_{\rho} ({\bf r}- {\bf r}^{\prime})
\label{2box1}
\end{eqnarray}
where the subscripts $V_1$ and $V_2$ denote that  ${\bf r}$ and
${\bf r}^{\prime}$ run over boxes 1 and 2 respectively.  Since these two
boxes are separated by a $\delta{\bf r}|$ that is much greater than the linear
extent of either box, ${\bf r}- {\bf r}^{\prime}$ is, to a good
approximation, equal to $|{\bf \delta r}|$ over the entire range of both integrals
in (\ref{2box1}).  I can therefore replace ${\bf r}- {\bf r}^{\prime}$
with
${\bf \delta r}$ in (\ref{2box1}), pull $C_{\rho} ({\bf \delta r})$ (which is now
independent of
${\bf r}$ and ${\bf r}^{\prime}$) out of the integrals, and perform the
integrals over ${\bf r}$ and ${\bf r}^{\prime}$.  Doing so gives
 \begin{eqnarray}
\left< \delta N_1 \delta N_2 \right> = C_{\rho} ({\bf \delta r}) \int _{V_1}  d^dr
\int _{V_2} d^dr^{\prime} = C_{\rho} ({\bf \delta r}) V_1V_2
\label{2box2}
\end{eqnarray}
where $V_{1,2}$ denote the volumes of boxes 1 and 2.  Rewriting  this in
terms of the mean numbers $\left< N_1\right>$ and  $\left< N_2\right>$ of
particles in each of the two boxes, and then using our earlier expression
(\ref{Crhorscale}) for $C_{\rho} ({\bf r})$, gives:
\begin{eqnarray}
 <\delta N_1 \delta N_2> = <N_1> <N_2> |{\bf \delta r}|^{-\alpha(d)}
G_d(\theta_{{\bf \delta r}})/\rho_0^2   .
\label{2box3}
\end{eqnarray}

Here $N_{1,2}$ are the particle numbers in box number 1 and 2,
respectively, and $\delta N_{1,2}$ are their fluctuations about their
mean.

Thus, this two box measurement  provides a direct measure of $\alpha
(d)$, which, as noted earlier, changes appreciably between two
and three dimensions. It also provides the opportunity to directly test my extremely detailed predictions (\ref{Gdef}), (\ref{Up2d}), and (\ref{Up3d})  for the functional form of $G_d(\theta_{{\bf r}})$.

\section{Summary}

I have used  hydrodynamic equations of dry active polar fluids\cite{TT1,TT2,TT3,TT4,NL}  to show that these systems exhibit giant number fluctuations, much larger the
``law of large numbers" scaling $\sqrt{<N>}$ scaling of number fluctuations in virtually all other 
equilibrium and non-equilibrium systems studied to date. Furthermore, I've shown that, again unlike  most other systems, 
the number fluctuations
also depend singularly on the shape of the box in which one counts the particles, vanishing in the limit of very thin boxes. 

These fluctuations arise {\it not} from large density fluctuations(which are not, in fact, expected in general in \dry s ), but from  long ranged
spatial correlations between those fluctuations. These can be determined by a surprising electrostatic analogy. Specifically, in two spatial dimensions the correlations can be obtained  by a simple rescaling of lengths from the electrostatic potential near a sharp upward pointing conducting wedge of opening angle ${3\pi\over8}=67.5^\circ$, while in three dimensions they can likewise be obtained in the same manner from  the electrostatic potential near a sharp upward pointing charged cone of opening angle $37.16^\circ$. This very precise prediction can be stringently tested  by alternative correlating number counts in two widely separated boxes.
% \end{abstract} 

\section*{Acknowledgements} I am very grateful to Francesco Ginelli for introducing me to the idea of Giant number fluctuations in \dry s, and to Sriram Ramaswamy for likewise introducing me to them in active nematics. I also thank the  Max Planck Institute for the Physics of Complex Systems,  Dresden; the Department of Bioengineering at Imperial College, London; The Higgs Centre for Theoretical Physics at the University of Edinburgh; and the Lorentz Center of Leiden University, for their hospitality while this work was underway.


\begin{thebibliography}{000}
\bibitem{BEC} Except in Bose-Einstein Condensates, in which  rms number
fluctuations scale like $\left< N\right>$ for temperatures $T < T_c$, where $T_c$ is the Bose-Einstein condensation temperature.
\bibitem{d>4} In fact, for \dry s in spatial dimensions $d>4$, $\phi=\frac{1}{2}+\frac{1}{d}$, which is also $>{1\over2}$ for {\it all} $d>4$ as well. I will not further discuss this obviously unphysical case here, except to note that it demonstrates that the Giant number fluctuations do {\it not} depend on a ``breakdown of linearized hydrodynamics" for their existence, since no such breakdown occurs in $d>4$ for \dry s, as shown in \cite{TT1, TT2, TT3, TT4, NL}.
\bibitem{foot0} An
estimate for which is given after eqn. 52 of reference \cite{TT4}.
\end{thebibliography}
\end{document}